\documentclass[prb,twocolumn,superscriptaddress,longbibliography]{revtex4-2}
 \usepackage{amsmath,amssymb, mathrsfs}
 \usepackage{stmaryrd}
 \usepackage{latexsym}
 \usepackage{graphicx}
 \usepackage{indentfirst}
 \usepackage{listings}
 \usepackage{xcolor}
 \usepackage{booktabs}
 \usepackage{stmaryrd}
 \usepackage{multirow}
 \usepackage{verbatim}
 \usepackage{makecell}
 \usepackage{lineno,hyperref}
 \usepackage{longtable}
 \usepackage{hyperref}
  \usepackage{lineno}
  \usepackage{soul}

\hypersetup{colorlinks,breaklinks,
            linkcolor=blue,urlcolor=blue,anchorcolor=blue,citecolor=blue}

\newcommand{\ud}{\mathrm{d}}

\newcommand{\txm}{\text{m}}
\newcommand{\txp}{\text{p}}

\newcommand{\txs}{\text{s}}

\newcommand{\uB}{\mathrm{B}}

\begin{document}

\title{Coherent and Semicoherent $\alpha/\beta$ Interfaces in Titanium:  \\structure, thermodynamics, migration}

\author{Siqi Wang}
\affiliation{Department of Materials Science and Engineering, City University of Hong Kong, Hong Kong SAR, China}
\author{Tongqi Wen}
\affiliation{Department of Mechanical Engineering, The University of Hong Kong, Pokfulam Road, Hong Kong SAR, China}
\author{Jian Han}
\email[Corresponding Author: ]{jianhan@cityu.edu.hk}
\affiliation{Department of Materials Science and Engineering, City University of Hong Kong, Hong Kong SAR, China}
\author{David J. Srolovitz}
\affiliation{Department of Mechanical Engineering, The University of Hong Kong, Pokfulam Road, Hong Kong SAR, China}

\date{\today}

\begin{abstract}
The $\alpha/\beta$ interface is central to the microstructure and mechanical properties of titanium alloys.
We investigate the structure, thermodynamics and migration of the coherent and semicoherent Ti $\alpha/\beta$ interfaces as a function of temperature and misfit strain via molecular dynamics (MD) simulations, thermodynamic integration and an accurate, DFT-trained Deep Potential.
The structure of an equilibrium semicoherent interface consists of an array of steps, an array of misfit dislocations, and coherent terraces.
Analysis determines the dislocation and step (disconnection) array structure and habit plane.
The MD simulations show the detailed interface morphology dictated by intersecting disconnection arrays.
The steps are shown to facilitate $\alpha/\beta$ interface migration, while the misfit dislocations lead to interface drag; the drag mechanism is different depending on the direction of interface migration.
These results are used to predict the nature of  $\alpha$ phase nucleation on cooling through the $\alpha$-$\beta$ phase transition.
\end{abstract}

\maketitle

\section{Introduction}

Ti alloys have received widespread attention for their superior mechanical properties, low density and biocompatibility over several decades.\cite{Ezugwu_1997,Peters_2003,GLutj1998}
The microstructure of Ti alloys is important due to its significant influence on  alloy mechanical performance~\cite{leyens2003titanium,Kundu_2012,Shang_2020}.
The microstructure of many Ti alloys used in structural applications is a mixture of $\alpha$ (hexagonal close-packed, HCP) and $\beta$ (body-centered cubic, BCC) phases.
$\alpha/\beta$ Ti alloys achieve a favorable balance between strength, ductility, fracture toughness and formability; this explains their widespread use in aerospace and other industries~\cite{leyens2003titanium,Jianwei_2019,SL2022}.
The two-phase nature of $\alpha/\beta$ Ti alloys implies the existence of several microstructural degrees of freedom that may be manipulated to achieve the desired mechanical property profile~\cite{semiatin2020overview}.
Since the microstructure represents a spatial distribution of $\alpha$/$\beta$ interfaces, understanding  the structure and thermodynamics of this interface is prerequisite to microstructure optimization.\cite{LiMei_2019,Filip_2003,Pengfei_2020} 

The central quantity for determining interfacial thermodynamics and kinetics is the interface (free) energy.
Interface energy determines the (near-) equilibrium interface morphology~\cite{herring1951some}.
The interface energy  is the main factor in the capillary driving force in microstructural evolution~\cite{taylor1992overview,Alan_1998}.
Interface energy is also required for the prediction of the barriers for precipitate nucleation.
Additionally, interface energy is a key ingredient in the theory of interface diffusion\cite{gupta2003diffusion}, faceting-defaceting\cite{Huanzhao_2020}, interface segregation\cite{Dregia_1991}, intergranular fracture\cite{Bo_2012}, etc~\cite{SuttonBalluffi,Alber_2010}.
Unfortunately, the determination of the $\alpha$/$\beta$ interface energy of Ti is not straightforward.
Li et al.~\cite{Li_2016} obtained the energy of a coherent $\alpha$/$\beta$ interface in Ti at 0~K via density functional theory (DFT) methods.
Since  $\beta$  is unstable at 0~K, they were unable to fully relax the $\beta$ structure without artificial constraints.
The $\alpha$ and $\beta$ pure Ti phases only coexist at finite temperature (without artificial constraint); hence, the interface free energy should be obtained at  finite temperature.
Unfortunately, it is impractical to directly determine the finite-temperature interface free energy via DFT.
Another important consideration is that most $\alpha$/$\beta$ interfaces observed in experiments are semicoherent, i.e., coherent interfaces decorated by misfit disconnections.
Calculation of semicoherent interface energy requires large-scale simulations which cannot be handled by DFT.
Interface energies are not easily determined from experiments either.
While Murzinova et al.~\cite{Murzinova_2016} estimated the $\alpha$/$\beta$ semicoherent interface energy based on the terrace-ledge model and linear elasticity using experimentally measured parameters, there is no direct experimental measurement of the $\alpha$/$\beta$ interface energy in Ti or Ti alloys.

Two recently developed techniques provide a path for us to determine the finite-temperature energy for the $\alpha$/$\beta$ semicoherent interface in Ti.
One is thermodynamic integration with the adiabatic switching free-energy calculation method~\cite{frenkel2001understanding,watanabe1990direct,de1996einstein,de1997adiabatic}.
This method is accurate (with fewer assumptions than the harmonic-approximation) and has proven efficient in determining the interface free energy~\cite{Freitas_2016}.
Another technique is to use a neural network potential trained with DFT data.
Here, we use the Deep Potential (DP) neural network potential~\cite{zhang2018deep} developed  by Wen et al.~\cite{wen2021specialising}.
In this paper, we apply both techniques to study the structure and energy of $\alpha$/$\beta$ coherent/semicoherent interfaces at finite temperatures.


This paper is organized as follows.

We first focus on the thermodynamic properties of the coherent $\alpha$/$\beta$ interface in titanium (i.e.,  $(0\bar{1}10)^\alpha \parallel (1\bar{1}2)^\beta$ and $[0001]^\alpha \parallel [110]^\beta$) as a function of strain and temperature.
Next, we examine the structure and properties of the  semicoherent interface. 
This information is then applied to understand the nucleation and growth of  $\alpha$ precipitates in a $\beta$ matrix (i.e., cooling from high temperature).
The main findings in this paper are as follows.  
(i) We predict the free energy of the most important interfaces  (coherent and semicoherent)  in titanium. This represents the first such calculations with DFT-level accuracy (note that $\beta$ phase is completely unstable at 0 K and hence inaccessible to  DFT without artificial constraints).
(ii) Our simulations show the equilibrium structure of the semicoherent interface and its intrinsic defect structure that gives rise to the widely-observed habit plane.
(iii) We demonstrate the mechanism of interface migration and that this mechanism gives rise to different interface mobilities in different directions (heating vs. cooling).
(iv) These accurate thermodynamic and structural results are applied to make reliable predictions on  how precipitation occurs upon cooling through the  $\alpha$-$\beta$ phase transition.  
This paper provides a roadmap for accurate prediction of interface properties and motion as well as precipitation in any system, including in systems with phases that are unstable at low temperature and in systems where loss of coherency occurs.

\section{Results}
\subsection{$\alpha$ and $\beta$ phases of titanium}\label{numerical}

To predict the properties of the $\alpha$ (HCP)/$\beta$ (BCC) interface in Ti, we initially determine the phase stability and bulk free energies of these phases.
We perform MD simulations to determine the lattice constants as functions of temperature and the free energies of the two phases.
This provides essential information on the stability and metastability of the two phases.
The perfect crystals are simulated using periodic boundary conditions in all directions and the simulation cell edge lengths and edge angles are free to change during the structural relaxation.


\begin{figure}[tb]
\begin{center}
\includegraphics[width = 0.95\linewidth]{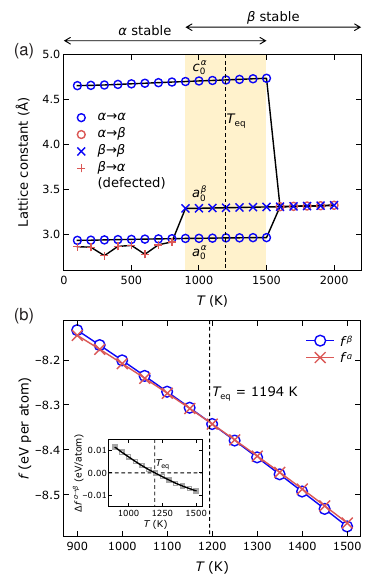}
\caption{Basic properties of $\alpha$ and $\beta$ phase}%
{(a)~Lattice constant vs. temperature for the initial $\alpha$ (HCP) phase and $\beta$ (BCC) phase.
The red (blue) symbols denote the cases where the structure changes (does not change) upon relaxation at different temperatures.
The yellow shaded region indicates the temperature range within which both  $\alpha$ and $\beta$  are stable/metastable.
The dashed line indicates the equilibrium temperature of the two phases $T_\text{eq}$ obtained from (b).
(b)~The stress-free free energies per atom vs. temperature for bulk $\alpha$ (red crosses) and  $\beta$ Ti (blue circles).
The inset shows the free energy difference $\Delta f^{\alpha\to\beta}$ ($\equiv f^\beta - f^\alpha$) vs. temperature.
}
\label{fig:latticeconstant}
\end{center}
\end{figure}

The single-phase $\alpha$ and $\beta$ Ti phases were equilibrated at different temperatures, where the size and shape of the simulation cell was fully relaxed under zero traction boundary conditions.
Figure~\ref{fig:latticeconstant}a shows the simulation results.
Each data point represents an independent simulation at a temperature.
Examination of the temperature dependence of lattice constants shows the temperature range where  $\alpha$ and $\beta$ are stable/metastable.
For the cases corresponding to the blue open circles (blue crosses), the simulation starts in the $\alpha$ ($\beta$) phase and the structure remains unchanged during thermal equilibration.
The red open circles (red pluses) indicate that $\alpha$ ($\beta$)  transforms  to perfect $\beta$ ($\alpha$)  upon finite temperature equilibration.
The red pluses imply that  $\beta$  transforms to defected $\alpha$  (containing many stacking faults).
The yellow shaded region shows where the two phases coexist (one stable, one metastable); i.e., $900$~K $\gtrsim T\gtrsim1500$~K.


The Gibbs phase rule implies that for a single-component (Ti) system, two phases ($\alpha$ and $\beta$) can coexist with one degree of freedom (temperature or pressure); i.e., two stress-free phases coexist at a particular temperature for each stress/pressure.
The coexistence temperature occurs where the free energies of the two phases are identical.
Figure~\ref{fig:latticeconstant}b shows the change in bulk free energy per atom with temperature for the $\alpha$ ($f^\alpha$) and $\beta$ ($f^\beta$) phases.
In general, the free energy of each phase decreases with increasing temperature (positive entropy).
The free energy curves cross at  $T_\text{eq} = 1194$~K -- this is the equilibrium temperature for a stress-free two-phase system (denoted by the vertical dashed lines in Figs.~\ref{fig:latticeconstant}a and~\ref{fig:latticeconstant}b.
Below $T_\text{eq}$, the free energy of  $\alpha$  is lower than that of  $\beta$; i.e., $\alpha$  is more stable than  $\beta$ for $T < T_\text{eq}$ ($\beta$ is metastable).
The inset of Fig.~\ref{fig:latticeconstant}b shows the free energy difference $\Delta f^{\alpha\to\beta}$ ($\equiv f^\beta - f^\alpha$) vs. temperature.
Expanding the free energy about $T = T_\text{eq}$ (to  first order) yields $\Delta f^{\alpha\to\beta}(T)
\approx ({m\ell}/{T_\text{eq}})(T_\text{eq} - T)$,
where $\ell$ is the specific latent heat.
From the inset of Fig.~\ref{fig:latticeconstant}b, we find the latent heat of the $\alpha \to \beta$ transition to be $75$~J~$\mathrm{g}^{-1}$, close to the experimental measurement, $90$~J~$\mathrm{g}^{-1}$~\cite{kaschnitz2002enthalpy}.

\subsection{Coherent $\alpha$/$\beta$ interface}\label{section:coherent}

A coherent interface can be constructed by matching the lattices of two phases along the interface plane with a small structural period.
Two lattices cannot typically be perfectly matched at their equilibrium lattice constants.
This implies that  one or both lattices must be strained to match along the interface plane.
These coherency strains necessarily increase the free energy of each of the strained phases. 
Although the coherent interface does not correspond to the most commonly observed habit plane, it is important because it has the  lowest energy among all possible $\alpha$/$\beta$ interfaces. 
When cooling titanium from the high-temperature $\beta$ phase, $\alpha$ phase particles nucleate and grow (see Sect.~\ref{section:discussion}).
The dominant orientation relationship is established in the nucleation stage of precipitation.
When the precipitate is small, interface energy dominates elastic energy and hence the lowest-energy interface occurs~\cite{SHI20124172}.
As the precipitate grows, elastic energy becomes increasingly important and the interface goes from coherent to semi-coherent.
The semicoherent interface consists of large terraces of coherent interface, separated by disconnections that have both dislocation character (relaxing the misfit) and step character (leading to a modest interface inclination from the Burgers/coherent interface relation).
We  investigate the semicoherent interface (on the commonly observed habit plane) in Sect.~\ref{section:semicoherent}.

Here, we focus on the free energy of the coherent $\alpha$/$\beta$ interface in Ti under different thermodynamic conditions.
We first address the crystallography, then investigate the variation of interface free energy with temperature at fixed coherency strain and the relationship between the interface free energy and the coherency strain.

\begin{figure}[tb]
\begin{center}
\includegraphics[width = 0.85\linewidth]{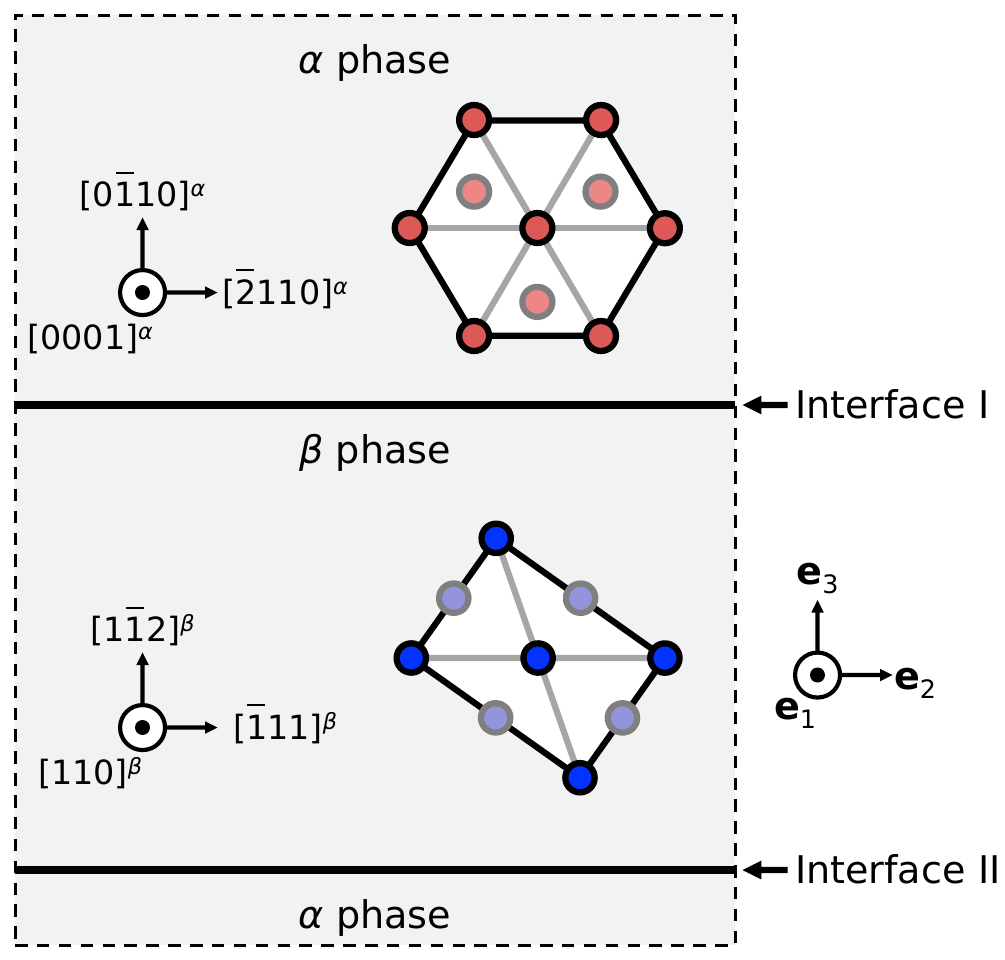}
\caption{Schematic plot for $\alpha$/$\beta$ interface}%
{Schematic of the simulation cell containing two identical coherent $\alpha$/$\beta$ interfaces.
  The atoms with darker/lighter color are located at different layers along the $\mathbf{e}_1$-axis.}
\label{fig:coherentinterface}
\end{center}
\end{figure}

Our two-phase  simulation model contains two identical coherent $\alpha$/$\beta$ interfaces; see Fig.~\ref{fig:coherentinterface}.
The simulation models contain 46,080 atoms in a single-phase $\alpha$ or $\beta$ system and 47,424 atoms in the two-phase/interface system.
The numbers of atoms are chosen as a trade-off between finite-size effects and  computational efficiency (the finite-size effects are examined in the Supplementary Information, SI).
The coordinate system is chosen such that the $\mathbf{e}_3$-axis is normal to the interface.
Periodic boundary conditions are applied in all directions (so, the system contains two identical interfaces), as shown in Fig.~\ref{fig:coherentinterface}.
The simulation cell is relaxed such that the stress  $\sigma_{33}=0$, the strain component $\epsilon^{\alpha/\beta}_{i3}=0$ ($i=1,2$)  and $\epsilon^{\alpha/\beta}_{ij}$ ($i,j=1,2$) are fixed at prescribed values during the simulations.

The commonly observed  $\alpha/\beta$ interface in Ti exhibits the Burgers orientation relationship (BOR): $(0\bar{1}10)^\alpha \parallel (1\bar{1}2)^\beta$ and $[0001]^\alpha \parallel [110]^\beta$, as shown in Fig.~\ref{fig:coherentinterface}.
The interface of BOR has lower interface energy than those of other candidate orientation relationships~\cite{Murzinova_2021,shi_2012,Da_2006}.
Given the equilibrium lattice constants of the bulk $\alpha$ and $\beta$ phases, perfect lattice matching in the BOR implies that  $\alpha$  must be compressed and/or  $\beta$  must be stretched along the $\mathbf{e}_1$- and $\mathbf{e}_2$-axes at all temperatures.

\begin{figure}[tb]
\begin{center}
\includegraphics[width = 0.95\linewidth]{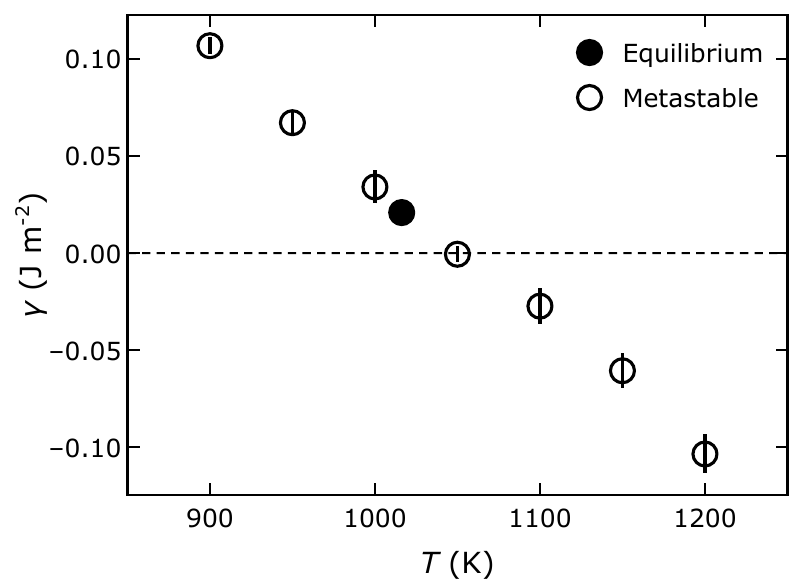}
\caption{Coherent interface free energy vs. temperature}%
{Coherent $\alpha/\beta$ interface free energy vs. temperature for the case in which the $\alpha$ crystal is strained to be coherent with the unstrained $\beta$.
Each error bar on each data point is obtained by six repeated computations.
The data point in black corresponds to the temperature at which the  $\alpha/\beta$ system is in equilibrium. }
\label{fig:fixedstrain}
\end{center}
\end{figure}

In our first set of interface simulations, we maintain the equilibrium lattice constant of  $\beta$  at the temperature of interest and compress  $\alpha$  in both the $\mathbf{e}_1$- and $\mathbf{e}_2$-directions to match  $\beta$  with the BOR; i.e., this corresponds to a scenario in which  an $\alpha$  lamella   grows from within  $\beta$.
We calculated the interface free energy at different temperatures by  $\lambda$ integration; 
see Sect.~\ref{theory} for the detailed calculation methods. 
The results are shown in Fig.~\ref{fig:fixedstrain}.
For each coherency strains, the $\alpha$/$\beta$ system is only in  equilibrium at one temperature (Gibbs phase rule).
For the case where  $\beta$  is at its own equilibrium lattice constant, the two-phase system is in equilibrium at $1016$~K (black solid circle in Fig.~\ref{fig:fixedstrain}).
This temperature is lower than $T_\text{eq}$ in Fig.~\ref{fig:latticeconstant}b because coherency strains in  $\alpha$ raise its free energy such that the red curve in Fig.~\ref{fig:latticeconstant}b shifts upwards and the intersection of two curves shifts towards the left (lower temperature).
The open circles in Fig.~\ref{fig:fixedstrain} correspond to the situation in which either the $\alpha$ or $\beta$ is metastable with respect to the other and the interface does not move within the simulation time.
The interface free energy can still be calculated for the metastable case by Eq.~\eqref{gammaA} although it is not thermodynamically well-defined.
In general, we find that the interface free energy decreases with increasing temperature.

\begin{figure*}[tb]
\begin{center}
\includegraphics[width=1\linewidth]{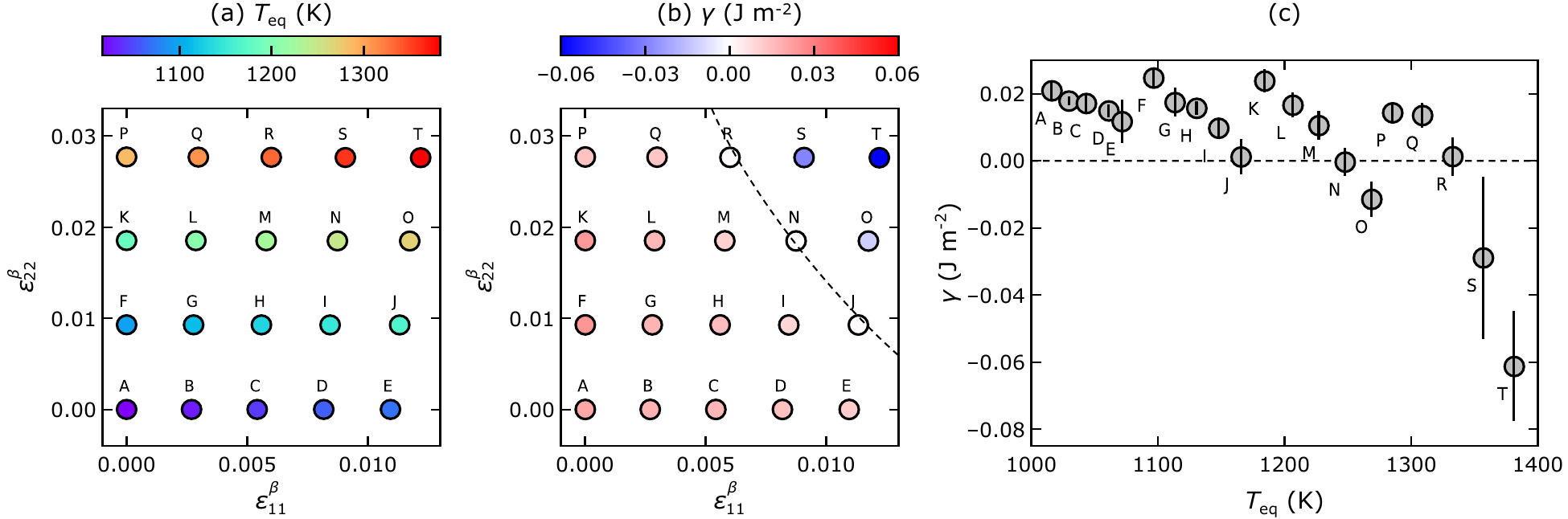}
\caption{Coherent interface free energy vs. temperature and strain}%
{(a) The equilibrium temperature $T_\text{eq}$ for a two-phase system containing a coherent interface and (b) the interface free energy $\gamma$ mapped on the strain space spanned by $\epsilon^\beta_{11}$ and $\epsilon^\beta_{22}$.
The dashed line in (b) approximately indicates the boundary between the region of positive $\gamma$ and the region of negative $\gamma$.
(c) The interface energy as a function of the equilibrium temperature; error bar is attached.
In all figures, the letters `A'-`T' labels are used to identify the same state.
}
\label{fig:efenergy}
\end{center}
\end{figure*}

The two phases in this unary system can be equilibrated along a coherent interface at different strains, corresponding to superimposing a biaxial strain on the system described above (Fig.~\ref{fig:fixedstrain}).
Each strain state has a unique equilibrium temperature.
Suppose that the structural periods of the coherent interface in the $\mathbf{e}_1$- and $\mathbf{e}_2$-directions are $p_1$ and $p_2$.
Since $\alpha$ and $\beta$ match with the BOR (Fig.~\ref{fig:coherentinterface}), $p_1$ is the length of $[110]^\beta a^\beta$ or equivalently $[0001]^\alpha c^\alpha$ and $p_2$ is the length of $[\bar{1}11]^\beta a^\beta/2$ or equivalently $[\bar{2}110]^\alpha a^\alpha/3$, where $a^\alpha$, $c^\alpha$ and $a^\beta$ are the lattice constants of  $\alpha$ and $\beta$  deformed in accordance with the coherency strain.
Then, the strains in $\beta$ parallel to the interface are
\begin{equation}
\epsilon^\beta_{11}
= \frac{p_1 - \sqrt{2}a^\beta_0}{\sqrt{2}a^\beta_0}
\quad\text{and}\quad
\epsilon^\beta_{22}
= \frac{p_2 - \sqrt{3} a^\beta_0/2}{\sqrt{3} a^\beta_0/2},
\end{equation}
where $a_0^\beta$ is the equilibrium lattice constant of $\beta$  at the temperature of interest.
We sampled the strain state within the range: $p_1\in [\sqrt{2}a_0^\beta, c_0^\alpha]$ and $p_2\in [\sqrt{3}a_0^\beta/2, a_0^\alpha]$.
The lower bound $(p_1, p_2) = (\sqrt{2}a_0^\beta, \sqrt{3}a_0^\beta/2)$ corresponds to the strain state $(\epsilon^\beta_{11}, \epsilon^\beta_{22}) = \mathbf{0}$, for which  $\beta$  is stress-free while  $\alpha$  is compressed to match  equilibrium $\beta$.
The upper bound $(p_1, p_2)=(c_0^\alpha, a_0^\alpha)$ corresponds to the case where  $\beta$  is stretched to match the equilibrium $\alpha$.
The cell size in the $\mathbf{e}_3$-direction is always fully relaxed.

For each strain state $(\epsilon^\beta_{11}, \epsilon^\beta_{22})$, we find the equilibrium temperature $T_\text{eq}$ from the intersection of the free energy-temperature curves of the two phases for different coherency strains (similar to Fig.~\ref{fig:latticeconstant}b).
The mapping of the equilibrium temperature on the strain space, $T_\text{eq}(\epsilon^\beta_{11}, \epsilon^\beta_{22})$, is shown in Fig.~\ref{fig:efenergy}a.
The variation in equilibrium temperature with strain  is $\lesssim400$~K; it is more sensitive to $\epsilon^\beta_{22}$ than $\epsilon^\beta_{11}$ (because  $\textbf{e}_2$ is the close-packed direction).
Based on the red ($f^\alpha(T)$) and blue ($f^\beta(T)$) curves in Fig.~\ref{fig:latticeconstant}b, we see that
when $\alpha$ is compressed to match $\beta$ (point `A' in Fig.~\ref{fig:efenergy}a), the $f^\alpha$-curve shifts upwards and $T_\text{eq}$ reduces to below 1194~K.
When $\beta$ is stretched to match $\alpha$ (point `T' in Fig.~\ref{fig:efenergy}a), the $f^\beta$-curve shifts upwards and $T_\text{eq}$ rises above 1194~K.

The interface free energy was calculated for each strain  (i.e., points `A'-`T' in Fig.~\ref{fig:efenergy}a) at the corresponding equilibrium temperature.
The interface free energy mapped on the strain space, $\gamma(\epsilon^\beta_{11}, \epsilon^\beta_{22})$, is shown in Fig.~\ref{fig:efenergy}b.
The  interface free energy decreases with increased stretch of  $\beta$  (decreased compression in  $\alpha$); and vice versa.
In the region above the dashed line in Fig.~\ref{fig:efenergy}b, the interface free energy is negative, indicating that the two-phase system is metastable.
This negative interface free energy is consistent with the metastability of $\beta$  at large tensile strains in the $\mathbf{e}_1$- and $\mathbf{e}_2$-directions.
This is evidenced by highly defected $\beta$ in the case where the interface free energy is negative; see Supplementary Information.

We replot the data in Fig.~\ref{fig:efenergy}b as the interface free energy vs. temperature in Fig.~\ref{fig:efenergy}c; points `A'-`T' correspond to the states labeled in Figs.~\ref{fig:efenergy}a and b.
We find that the data points are clustered into four groups: `A'-`E', `F'-`J', `K'-`O' and `P'-`T'.
Each group corresponds to the same $\epsilon^\beta_{22}$ and varying $\epsilon^\beta_{11}$ ($\epsilon^\beta_{11}$ is varied over a smaller range than   $\epsilon^\beta_{22}$).
There is a clear trend that the interface free energy $\gamma(\epsilon^\beta_{11}, \epsilon^\beta_{22})$, increases with decreasing $T_\text{eq}(\epsilon^\beta_{11}, \epsilon^\beta_{22})$.

The $\alpha$/$\beta$ coherent interface can exist over a range of strains and its energy is a function of those strains.
Along a semicoherent interface (i.e., a coherent interface with widely spaced dislocations/disconnections), the strain state varies with position along the coherent terraces (i.e., position along the terrace relative to the positions of the dislocations/disconnections).
If we know the strain distribution along a semicoherent interface (e.g., from continuum elasticity), we may write the interface energy as
\begin{equation}
\gamma=\frac{1}{A}\iint_A\gamma\left(\epsilon^\beta_{11}(x_1,x_2), \epsilon^\beta_{22}(x_1,x_2)\right)dx_1 dx_2,
\end{equation}
where $\left(\epsilon^\beta_{11}(x_1,x_2), \epsilon^\beta_{22}(x_1,x_2)\right)$ is from Fig.~\ref{fig:efenergy}b.
Of course, there are corrections for strain gradients.
An alternative approach is to simply do molecular dynamics simulations on a semicoherent interface.

\subsection{Semicoherent $\alpha$/$\beta$ interface}\label{section:semicoherent}

As the size of $\alpha$ or $\beta$ phase grows, the strain energy in the two-phase system becomes too large to remain coherent.
At this point, the misfit strains can be relaxed by introducing disconnections along the interface with finite Burgers vector components parallel to the interface plane.
In this section, we investigate the structure and energetics of the $\alpha$/$\beta$ semicoherent interface.

\begin{figure}[t]
\begin{center}
\includegraphics[width = 1\linewidth]{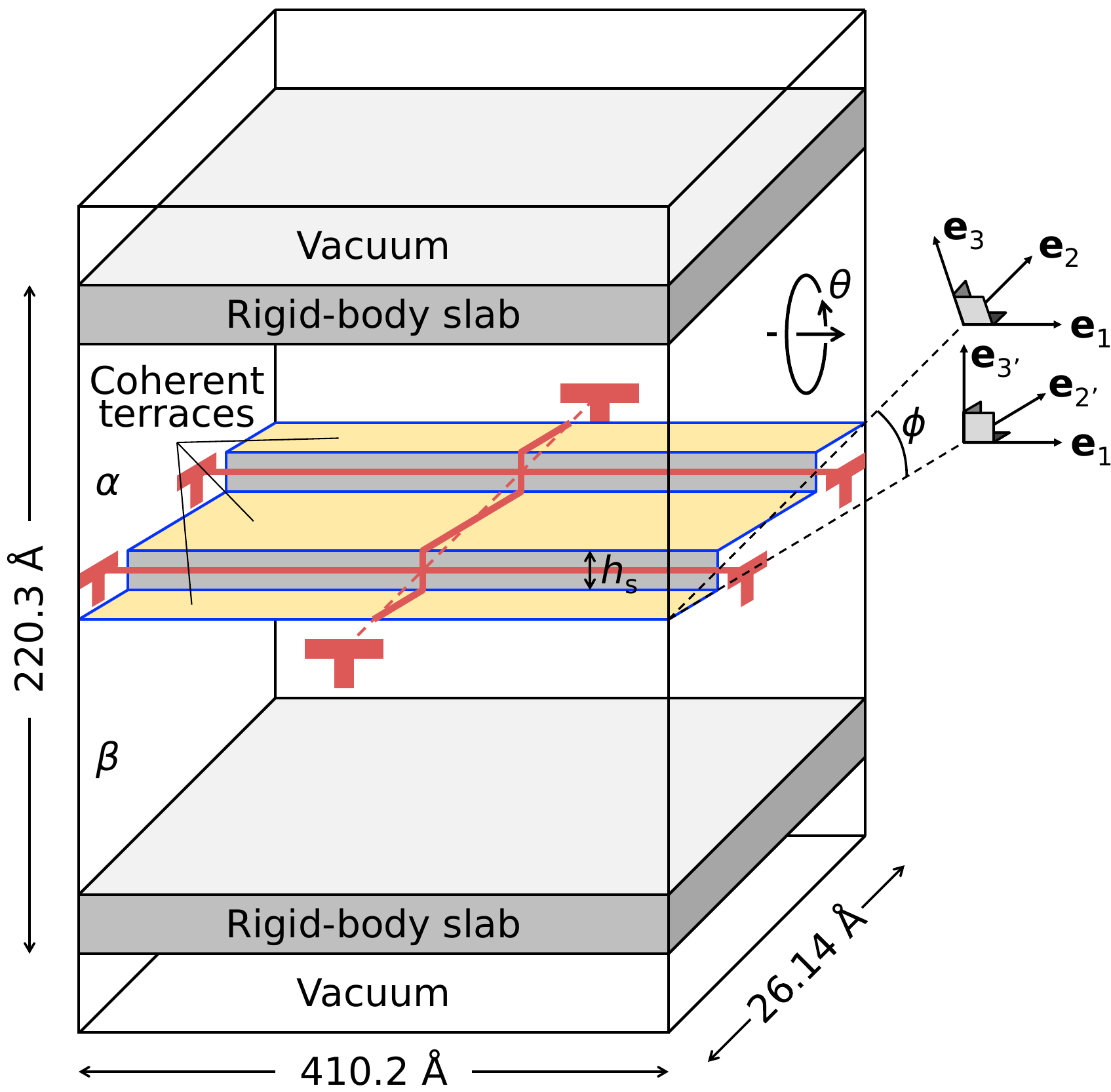}
\caption{Schematic of the $\alpha$/$\beta$ semicoherent interface model.}%
{The coordinate system $(\mathbf{e}_1,\mathbf{e}_2,\mathbf{e}_3)$ is attached to the mean interface plane (habit plane); $\mathbf{e}_3$ is the habit plane normal and $\mathbf{e}_1$ and $\mathbf{e}_2$ are parallel to the habit plane.
The coordinate system $(\mathbf{e}_1,\mathbf{e}_{2'},\mathbf{e}_{3'})$ is attached to the coherent interface plane.
The blue lines depict the interface profile; the yellow regions denote the coherent interface terraces.
The red lines parallel to the $\mathbf{e}_1$-axis represent the disconnections with step height $h_\txs$; the red line parallel to the $\mathbf{e}_2$-axis is a misfit dislocation with zero step height.
$\phi$ is the inclination angle of the interface plane (habit plane) with respect to the coherent interface plane.
$\theta$ is the rotation angle of  $\alpha$  about the $\mathbf{e}_1$-axis with respect to  $\alpha$  in the coherent interface model (i.e., the red lattice in Fig.~\ref{fig:coherentinterface}).
}
\label{fig:semiinterface}
\end{center}
\end{figure}

The simulation model of a semicoherent interface is shown in Fig.~\ref{fig:semiinterface}.
According to the phenomenological theory of martensite crystallography~\cite{wechsler1953theory,bowles1954crystallography,ackerman2020interface}, the semicoherent $\alpha$/$\beta$ interface plane should be a habit plane determined by geometry, as follows.
Take the coherent interface with the BOR as a reference (i.e., the misorientation angle $\theta$ and the inclination angle $\phi$ of the coherent interface are defined to be zero).
Based on the equilibrium lattice constants at the equilibrium temperature $T = 1194$~K  (see Fig.~\ref{fig:latticeconstant}), a simple calculation~(see Supplementary Information for details) shows that the inclination angle of the habit plane is $\phi \approx 10.9^\circ$ and the misorientation angle is $\theta \approx 0.523^\circ$ (Fig.~\ref{fig:semiinterface}); we construct the simulation cell with $\phi$ and $\theta$ close to these values.
The Cartesian coordinate system is established such that $\mathbf{e}_1 \parallel [110]^\beta$, $\mathbf{e}_3$ is perpendicular to the habit plane and $\mathbf{e}_2 = \mathbf{e}_3 \times \mathbf{e}_1$.
In practice, we choose the inclination angle
$\phi = \arctan\left[\left|[1\bar{1}2]^\beta a^\beta_0/3\right| \middle/ \left(5  \left|[\bar{1}11]^\beta a^\beta_0/2\right|\right)\right] \approx 10.7^\circ$
such that the periodic boundary condition is satisfied along the $\mathbf{e}_2$-axis. 
The simulation cell size is shown in Fig.~\ref{fig:semiinterface}.
The total number of Ti atoms in the simulation model is 91,560.

The surface layers are treated as rigid-body slabs with a thickness of $10$~\AA.
The relative coordinates of the atoms inside the surface slabs are fixed with the equilibrium lattice constants at the temperature of interest.
The surface slabs are allowed to relax \textit{en bloc} such that $\sigma_{i3} = 0$ ($i=1,2,3$).
The motion of atoms in the surface slabs is excluded from the $\lambda$ integration.

\begin{figure*}[tb]
\begin{center}
\includegraphics[width = 1\linewidth]{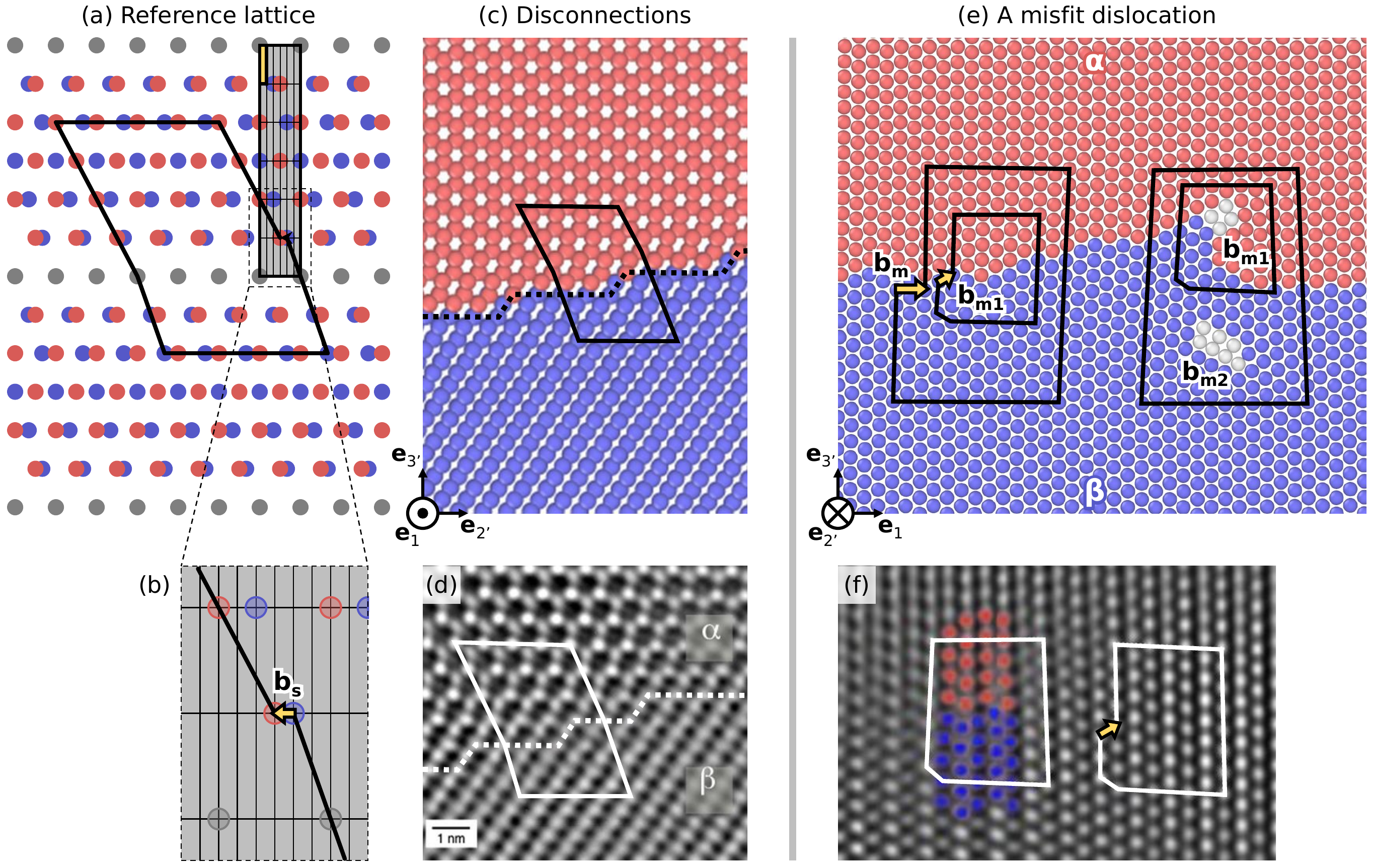}
\caption{Semicoherent interface structure}%
{(a) Reference dichromatic pattern formed by the BCC/$\beta$ lattice (blue) and HCP/$\alpha$ lattice (red) for the BOR.
The gray points denote the overlapped blue/red points.
The gray-shaded region is a CSL unit cell; the narrow yellow-shaded region is a DSC unit cell.
A Burgers circuit showing the closure failure is drawn; it corresponds to the closed circuit in (c).
(b) Enlargement of the region framed by dashed lines in (a), showing the DSC lattice and the Burgers vector (denoted by a yellow arrow).
(c) and (d) show, respectively, the simulation result and TEM image of $\alpha$/$\beta$ interface structure viewed along the $-\mathbf{e}_1$-axis, while (e) and (f) show the views along the $\mathbf{e}_2$-axis.
In (c) and (e), the red and blue atoms (colored according to the common neighbor analysis, CNA~\cite{1987cna}) denote $\alpha$ and $\beta$  phases respectively.
(d) and (f) are TEM images reproduced with permission~\cite{Zheng_2018} (Copyright 2018 Elsvier).
}
\label{fig:semicoherent_structure}
\end{center}
\end{figure*}


The equilibrium $\alpha$/$\beta$ semicoherent interface structure obtained from our simulations is shown in Figs.~\ref{fig:semicoherent_structure}c and e.
From the $\mathbf{e}_{2'}$-$\mathbf{e}_{3'}$ projection (Fig.~\ref{fig:semicoherent_structure}c), we find that the interface inclination is formed by superimposing a set of ``steps'' along the $\mathbf{e}_1$-axis on the coherent interface.
The characters of the ``steps'' can be deduced based upon the dichromatic pattern (see  Fig.~\ref{fig:semicoherent_structure}a).
The dichromatic pattern in Fig.~\ref{fig:semicoherent_structure}a is formed by interpenetrating the BCC  (blue) and  HCP lattices (red).
Since we use the BCC lattice ($\beta$) as our reference, the coherency strain is applied on the HCP lattice such that it matches the BCC lattice at the gray points.
The gray points form the coincidence-site lattice (CSL).
A CSL unit cell is shaded gray in Fig.~\ref{fig:semicoherent_structure}a.
In the CSL unit cell, the fine grid indicates the displacement-shift-complete (DSC) lattice.
The shift of the whole HCP lattice with respect to the BCC lattice by any DSC lattice vector preserves the dichromatic pattern.
Following the FS/RH convention~\cite{hirth1967theory}, we find that each interface step has an associated Burgers vector $\mathbf{b}_\txs = [\bar{1}11]a^\beta_0/12$, corresponding to a DSC lattice vector; see the yellow arrow Fig.~\ref{fig:semicoherent_structure}b.
Based on the dichromatic pattern in Fig.~\ref{fig:semicoherent_structure}a, a shift of the HCP lattice with respect to the BCC lattice by $\mathbf{b}_\txs$ necessarily results in the coincident sites (gray points) at the layer above the initial coincidence-site layer.
This suggests that $\mathbf{b}_\txs$ is associated with a step height $h_\txs = 2\sqrt{6}a_0^\beta/3$.
Hence, the ``steps'' observed in Fig.~\ref{fig:semicoherent_structure}c are disconnections characterized by Burgers vector $\mathbf{b}_\txs$ and step height $h_\txs$.
To distinguish this set of disconnections from another set of disconnections which will be discussed later, we refer to this set of disconnections as ``steps'' below (emphasizing the feature that $h_\txs \gg |\mathbf{b}_\txs|$).
The steps on the $\alpha$/$\beta$ interface are seen experimentally~\cite{Zheng_2018}, as shown in Fig.~\ref{fig:semicoherent_structure}d. 
The ideal step spacing (based on the topological model~\cite{Pond_2003}) along $\mathbf{e}_2$-axis is 14.34~{\AA} (see Supplementary Information for details).
We choose simulation cell dimension along the $\mathbf{e}_2$-axis that allows us to get close to this ideal value while remaining sufficiently small to be computationally tractable.
The step spacing in our simulation is 13.07~{\AA}.

\begin{figure*}[tb]
\begin{center}
\includegraphics[width = 1\linewidth]{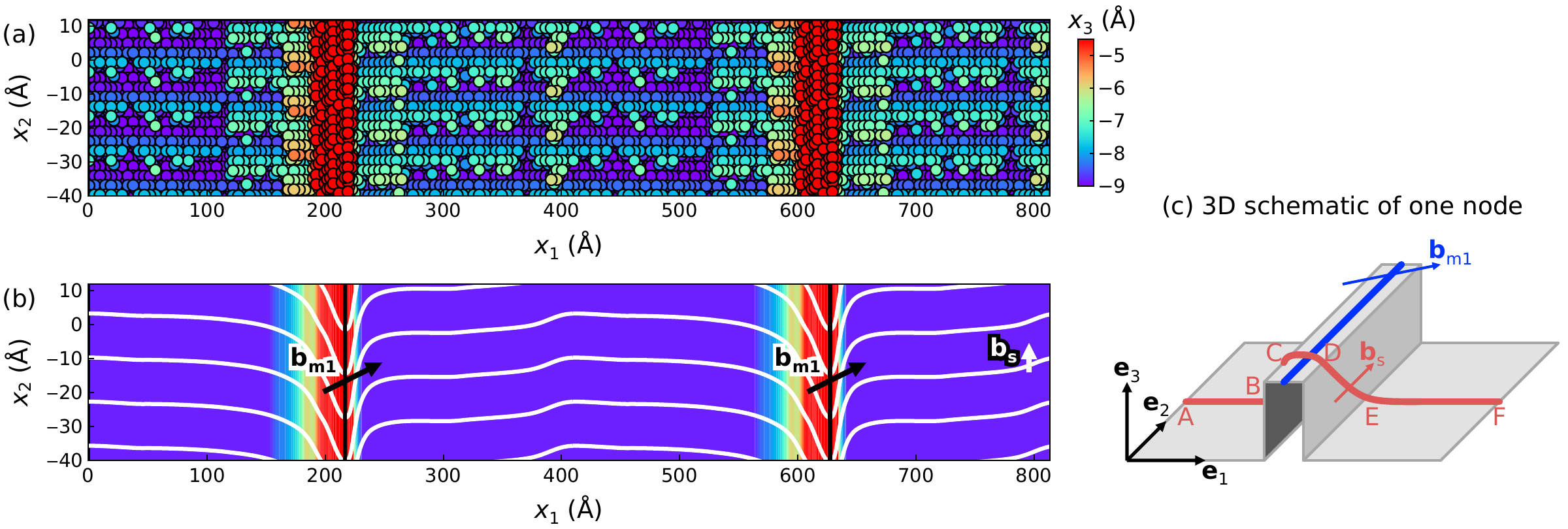}
\caption{Semicoherent interface dislocation structure}%
{(a) Plan view of the interface plane along the $-\mathbf{e}_3$-direction (the atoms in $\alpha$  are not shown ).
The atoms are colored according to their $x_3$ coordinates.
(b) The interface structure extracted from (a).
The color indicates the profile of the interface plane.
Two red regions indicate the interface plane humps.
The black lines denote BCC dislocations $\mathbf{b}_{\txm 1}$.
The white lines denote the steps with Burgers vector $\mathbf{b}_\txs$.
(c) 3D schematic of a   step $\mathbf{b}_\txs$ (red curve) and BCC dislocation $\mathbf{b}_{\txm 1}$ (blue curve) node.
The interface hump is depicted as  rectangular.
}
\label{fig:interfacestructure}
\end{center}
\end{figure*}

Figure~\ref{fig:semicoherent_structure}e shows the equilibrium semicoherent interface structure viewed along the $\mathbf{e}_{2'}$-direction.
The interface is composed of coherent sections and two dislocation lines directed along the $\mathbf{e}_2$-axis, labeled  $\mathbf{b}_{\txm 1}$ and $\mathbf{b}_{\txm 2}$.
From the dichromatic pattern in Fig.~\ref{fig:semicoherent_structure}a, the shortest DSC lattice vector parallel to the $\mathbf{e}_1$-axis is $[110]a_0^\beta$.
The Burgers vector of this disconnection, if it exists, would be $\mathbf{b}_\txm = [110]a_0^\beta$.
Since the relative shift of two lattices by $\mathbf{b}_\txm$ does not change the coincidence-site layer (the layer formed by the gray points), this disconnection is associated with zero step height; to distinguish it from the ``step'' disconnections discussed above, we call these ``misfit dislocations'' (zero step height disconnections) below. 
The theoretical value of the misfit dislocation spacing needed to relax the misfit strain is 412.75~{\AA}(see Supplementary Information).
We aligned 87 cells of $\alpha$ with 88 cells of $\beta$, resulting in a compressive  strain in $\beta$ of $\sim 0.07\%$ and a misfit dislocation spacing of 410.2~{\AA}.
Note that $|\mathbf{b}_\txm|$ is large.
According to the Frank energy criterion~\cite{frank1949discussion,hirth1967theory}, the misfit dislocation should undergo dissociation:
\begin{equation}
\begin{array}{ccccc}
\mathbf{b}_\txm &
\to&
\mathbf{b}_{\txm 1}&
+&
\mathbf{b}_{\txm 2} \\
\displaystyle{
\left[110\right] a_0^\beta } &
\to&
\displaystyle{
\left[111\right]\frac{a_0^\beta}{2} } &
+&
\displaystyle{
\left[11\bar{1}\right]\frac{a_0^\beta}{2} }
\end{array},
\end{equation}
where $\mathbf{b}_{\txm 1}$ and $\mathbf{b}_{\txm 2}$ are the Burgers vectors of two full dislocations in the BCC ($\beta$) lattice; we refer to these as ``BCC dislocations''.
$\mathbf{b}_{\txm 1}$ and $\mathbf{b}_{\txm 2}$ are the two dislocations observed in Fig.~\ref{fig:semicoherent_structure}e.
The Burgers vectors can be confirmed by drawing Burgers circuits with the FS/RH convention, as shown in Fig.~\ref{fig:semicoherent_structure}e.
By examining the Peach-Koehler force between the two BCC dislocations, we see that they are elastically attracted to one another (see Supplementary Information), such that the separation between them is finite.

Examination of the transmission electron microscopy (TEM) image in Fig.~\ref{fig:semicoherent_structure}f~\cite{Zheng_2018} shows only the $\mathbf{b}_{\txm 1}$ dislocation (confirmed by drawing a Burgers circuit).
While the complementary dislocation $\mathbf{b}_{\txm 2}$ must exist to accommodate the misfit between $\alpha$ and $\beta$, it is not observed.
Perhaps it is out of the field of view of the TEM image or it glides out of the TEM foil.
We also note from Fig.~\ref{fig:semicoherent_structure}e that the interface plane is locally curved towards $\alpha$ and terminates at a BCC dislocation $\mathbf{b}_{\txm 1}$.
Indeed, Ackerman et al.~\cite{ackerman2020interface}  experimentally observed widely spaced ``bumps'' on the interface  when the interface was imaged along  $[\bar{2}110]^\alpha/[\bar{1}11]^\beta$ direction.

We can understand the existence of the observed interface profile as follows.
To accommodate the mismatch in the $\mathbf{e}_1$-direction, the two BCC dislocations (from the dissociation of a misfit dislocation), should be located symmetrically about the mean interface plane.
Analysis of the interaction between the two BCC dislocations and the steps (with a small Burgers vector $\mathbf{b}_\txs$) also suggests that the interface which contains a set of steps should be located with equal distance to the two BCC dislocations (see Supplementary Information).
Most of the interface plane is indeed located between the two BCC dislocations.
However, the interface plane near the misfit dislocation arches towards $\alpha$  to contact one of the BCC dislocations $\mathbf{b}_{\txm 1}$, as shown in Figs.~\ref{fig:semicoherent_structure}e and \ref{fig:interfacestructure}a.
The arch of the interface can be understood as follows.
As explained previously, the interface plane should be located between the two BCC dislocations.
If so, $\mathbf{b}_{\txm 1}$ would sit inside  $\alpha$ .
However, $\mathbf{b}_{\txm 1}$ is a full BCC ($\beta$) lattice dislocation which cannot exist within HCP ($\alpha$) lattice.
Hence, the interface plane has to be curved to guarantee that $\mathbf{b}_{\txm 1}$ remains within  $\beta$.

The shape of the steps can be extracted from the $x_3$-coordinates of the atoms on the $\beta$-phase side of the interface, as shown in Fig.~\ref{fig:interfacestructure}a.
Figure~\ref{fig:interfacestructure}b shows the profile of the interface plane (colored contours) and the shape of the step lines (white curves) projected on the $\mathbf{e}_1$-$\mathbf{e}_2$ plane (habit plane).
Figure~\ref{fig:interfacestructure}c shows a simplified schematic of the node between a step $\mathbf{b}_\txs$ (red curve) and a BCC dislocation $\mathbf{b}_{\txm 1}$ (blue curve).
The step  does not simply rest on the interface hump with the shortest length, but bows slightly  in the $-\mathbf{e}_2$-direction.
This step bowing is energetically favorable.
Based on the simplified model shown in Fig.~\ref{fig:interfacestructure}c,
the segments of the step line, `BC' and `DE', lie on the side faces of the interface hump.
`BC' and `DE' are associated with the dislocation (Burgers vector $\mathbf{b}_\txs$, zero step height).
The tilting of segments `BC' and `DE' increases the segment length (compared with  untilted segments) and the line energy, while simultaneously reducing the elastic energy since the tilt-induced screw components have lower energy than edges.
The finite tilt angle of segments `BC' and `DE' is a consequence of the trade-off between the two factors -- the increase of segment length (raising the energy) and the increase of screw component (lowering the energy).

\begin{figure}[t]
\begin{center}
\includegraphics[width = 1\linewidth]{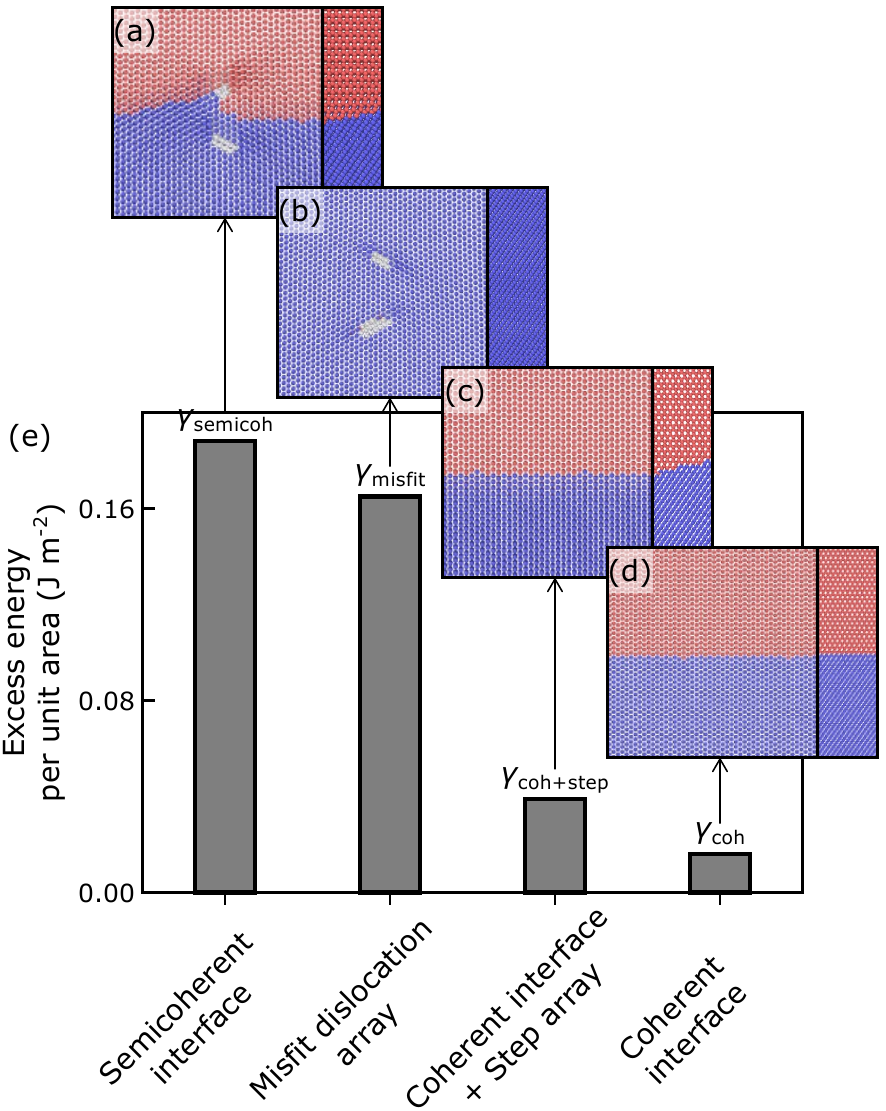}
\caption{Comparison of the semicoherent interface energy}%
{The  misfit dislocation energy and the coherent interface energy (see Section~\ref{section:coherent}).}
\label{fig:dislocationenergy}
\end{center}
\end{figure}

\label{eqsemicohgamma}

To obtain the equilibrium interface free energy, we construct five atomic models: perfect $\alpha$ bulk, perfect $\beta$ bulk, $\alpha$  with two surfaces, $\beta$  with two surfaces, and a two-phase system with two surfaces (here, ``surface'' always refers to the rigid-body surface slab).
The semicoherent interface free energy $\gamma$ is
\begin{equation}
\gamma A = F - N f - \gamma_{\txs\alpha} A - \gamma_{\txs\beta} A,
\end{equation}
where $F$, $N$ and $A$ are the total free energy, the number of atoms and the interface area of the two-phase system, $f$ is the free energy per atom of perfect $\alpha$ or $\beta$ bulk (in equilibrium $f^\alpha = f^\beta = f$), and $\gamma_{\txs\alpha/\txs\beta}$ is the excess energy per unit area due to the presence of rigid-body surfaces for $\alpha$/$\beta$, respectively.
$\gamma_{\txs\alpha/\txs\beta}$ is obtained by
\begin{equation}
\gamma_{\txs\alpha/\txs\beta} A_{\txs\alpha/\txs\beta}
= \frac{1}{2}
\left(F_{\txs\alpha/\txs\beta} - N_{\txs\alpha/\txs\beta} f\right),
\end{equation}
where $F_{\txs\alpha/\txs\beta}$, $N_{\txs\alpha/\txs\beta}$ and $A_{\txs\alpha/\txs\beta}$ are the total free energy, the number of atoms and the area of the $\alpha$/$\beta$ interface.
Based on this approach, we find that the semicoherent interface free energy is $0.188$~J~m$^{-2}$ at $T$ = 1194~K.

The semicoherent interface free energy can be partitioned into contributions from
(1) the $\alpha$/$\beta$ coherent interface,
(2) a periodic array of misfit dislocations (each  consisting of two BCC dislocations, $\mathbf{b}_{\txm 1}$ and $\mathbf{b}_{\txm 2}$),
(3) a periodic array of steps (disconnections with Burgers vector $\mathbf{b}_\txs$ and step height $h_\txs$),
(4) the nodes between  steps and BCC dislocations $\mathbf{b}_{\txm 1}$, and
(5) the elastic interaction between the misfit dislocation array and the step array.
We estimate these contributions.
To validate our calculation result for the semicoherent interface free energy $\gamma_\text{semicoh}$, we check to see if $\gamma_\text{semicoh}$ approximately equals the sum of the energies due to the above contributions at $1194$~K.
The $\alpha$/$\beta$ coherent interface free energy $\gamma_\text{coh}$ has been reported in Section~\ref{section:coherent}.
The free energy of an array of misfit dislocations (i.e., a pair of BCC dislocations) cannot be obtained without the introduction of a coherent interface.
As a crude approximation, we calculated the free energy of the same set of BCC dislocations in a bulk $\beta$ crystal with the same geometry as that of the two-phase system, $\gamma_\text{misfit}$; see Fig.~\ref{fig:dislocationenergy}b (see the simulation details in SI).
We also calculated the free energy of the configuration consisting of a coherent interface and an array of steps, $\gamma_\text{coh+step}$; see Fig.~\ref{fig:dislocationenergy}c.

Figure~\ref{fig:dislocationenergy}e shows the contributions to the interface free energy.
First, $\gamma_\text{coh+step}$ is almost twice of $\gamma_\text{coh}$; the difference is associated with the step array.
Second, $\gamma_\text{semicoh}$ is close to $\gamma_\text{misfit} + \gamma_\text{coh+step}$; this is reasonable if we assume that the contributions of (4) and (5) above are negligible.
Third, we note that $\gamma_\text{semicoh}$ is dominated by $\gamma_\text{misfit}$.
From the perspective of thermodynamics, the major difficulty for the formation of semicoherent interfaces is the introduction of the misfit dislocation array.
Banerjee et al.~\cite{banerjee2013perspectives} experimentally observed that the misfit dislocations come from the absorption of lattice dislocations from the matrix into a coherent interface, rather than from  atomic relaxation along the interface.

\section{Discussion}\label{section:discussion}

In this section, we discuss several implications of our simulation results.
Considering the fact that the interatomic potential employed in the simulations (i.e., the DFT-trained, Deep Potential) successfully reproduces many properties of Ti (including the phase diagram, crystal structures, defect properties, \ldots) \cite{wen2021specialising}, we have more confidence in the resultant predictions than for typical MD simulations.


\begin{figure*}[t]
\begin{center}
\includegraphics[width = 1\linewidth]{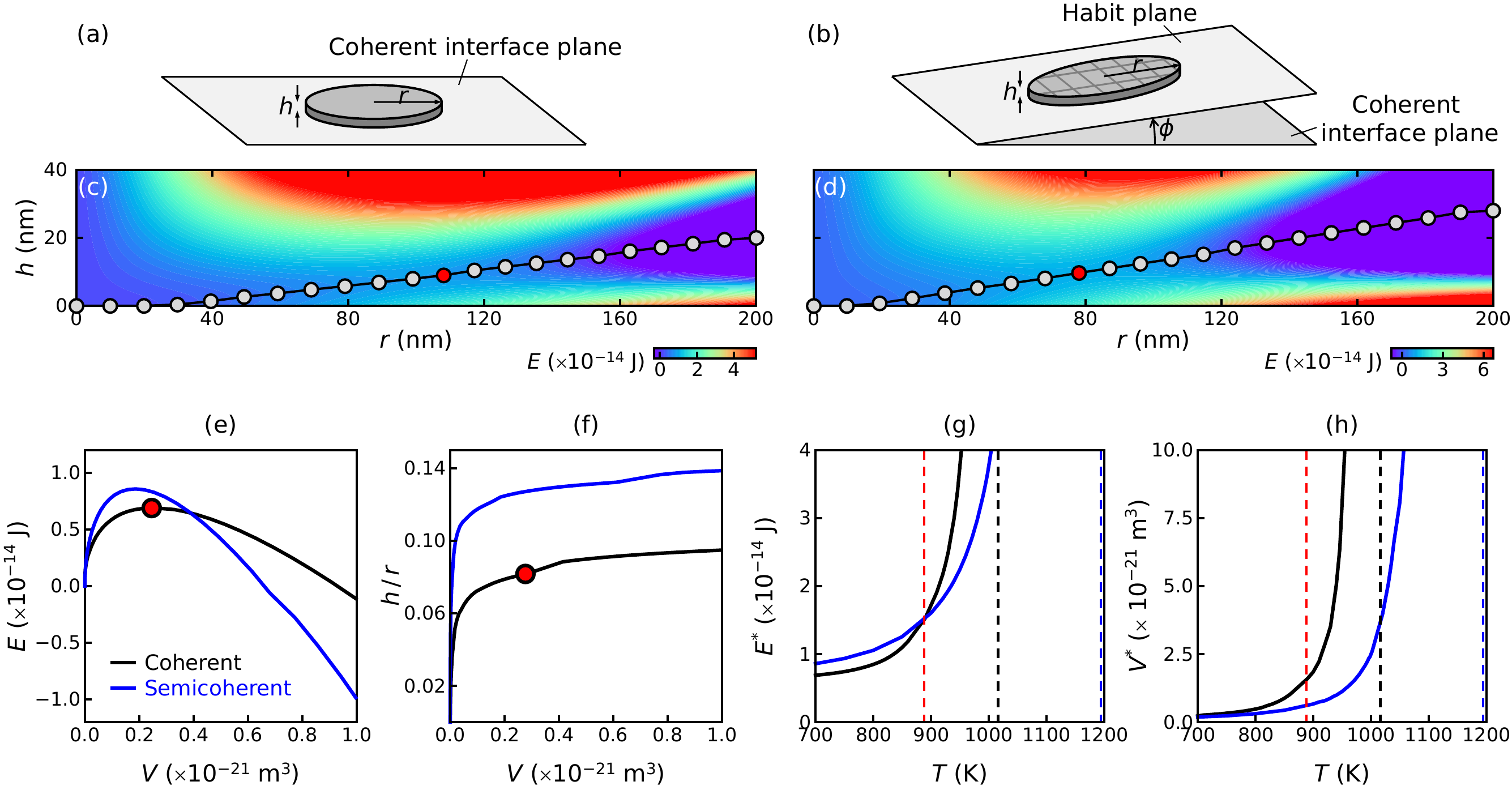}
\caption{Comparison of the nucleation path for coherent and semicoherent interface}%
{Schematics of the $\alpha$-phase nuclei with (a) coherent and (b)semicoherent interfaces.
The total free energy $E$ mapped to the space spanned by the nucleus radius $r$ and the thickness $h$ at $T=700$~K for the cases of (c) coherent and (d) semicoherent nuclei.
The points denote the states along the MEP obtained by NEB; the red points represent saddles.
(e) The energy and (f) the $h/r$ ratio vs. the inclusion volume $V$ along the MEP.
The red points indicate the values of $E$ and $h/r$ at the saddle point along the MEP for the growth of coherent nucleus, corresponding to the red point in (c).
(g) The energy barrier $E^*$ and (h) the critical nucleus volume $V^*$ vs. temperature.
The black and blue dashed lines indicate the thermodynamic phase transition temperatures for coherent and semicoherent nuclei, respectively.
The red dashed lines indicate  transitions between the critical coherent nucleus and the critical semicoherent nucleus.
}
\label{fig:energybarrier}
\end{center}
\end{figure*}

We first employ the present results to consider the early stages of nucleation of an $\alpha$ precipitate within a $\beta$ matrix upon cooling Ti through the $\alpha$/$\beta$ transition.
We focus on estimating the nucleation barrier and critical nucleus size based upon a straightforward model.
Experimental observations~\cite{banerjee2013perspectives,sharma2021fine} and phase-field simulations~\cite{shi2012predicting} suggest that the $\alpha$ nucleus is roughly a thin elliptic plate.
We approximate the $\alpha$ nucleus shape as a thin disk, as shown in Fig.~\ref{fig:energybarrier}, where $r$ and $h$ are the disk radius and thickness.
The flat surfaces of the plate may be $(0\bar{1}10)^\alpha$/$(1\bar{1}2)^\beta$ coherent interfaces (Fig.~\ref{fig:energybarrier}a) or semicoherent interface along the habit plane (Fig.~\ref{fig:energybarrier}b), for which the temperature-dependent coherent/semicoherent interface free energy is $\gamma(T)$ -- as calculated above.
The side surface of the plate may be approximated as a $(\bar{2}110)^\alpha$/$(\bar{1}11)^\beta$ interface with free energy $\gamma_\text{side}(T)$; see Supplementary Information.
The total energy of this plate-like nucleus is
\begin{align}\label{ErhT}
E(r,h;T)
&= \pi r^2 h \left[\Delta f^{\beta\to\alpha}(T) + f^\text{el}(T)\right]
\nonumber\\
&+ 2\pi r^2 \gamma(T)+ 2\pi r h \gamma_\text{side}(T),
\end{align}
where $\Delta f^{\beta\to\alpha} \equiv f^\alpha - f^\beta$ ($f^{\alpha/\beta}$ is the bulk free energy per unit volume of $\alpha$/$\beta$) and $f^\text{el}$ is the elastic inclusion energy (which also depends on temperature via lattice constant and elastic constant).
According to the mismatch between $\alpha$ and $\beta$, we evaluate $f^\text{el}$ using the Eshelby's inclusion method (see Supplementary Information).
The energies for the coherent and semicoherent nuclei at $T=700$~K are shown in Figs.~\ref{fig:energybarrier}a and b as functions of $r$ and $h$.
Based on the energy landscape $E(r,h; 700~\text{K})$, we searched the minimum energy path (MEP) and the saddle point by the free-end nudged-elastic-band method (FE-NEB~\cite{henkelman2000climbing,zhu2007interfacial}; see Supplementary Information). The energy and aspect ratio $h/r$ vs. nucleus volume $V$ along the MEP are shown in Figs.~\ref{fig:energybarrier}c and d.

At $T=700$~K, the nucleation barrier is lower for the coherent plate than the semicoherent plate.
For either the coherent or semicoherent plate, $h/r<0.15$, validating the assumption that the nucleus is a thin plate.
When the nucleus volume is larger than $\sim 0.4\times 10^6$~nm$^3$, the energy of the semicoherent plate is lower than that of the coherent plate, implying that as the initially coherent plate grows, interface coherency will be lost.
After the loss of coherency, the flat surfaces of the plate will be oriented along the habit plane and the aspect ratio will become larger.
The crossover in the energies for coherent and semicoherent nuclei and the resultant loss of coherency were seen earlier in phase-field simulations~\cite{shi2012predicting}.

Using the FE-NEB method and Eq.~\eqref{ErhT}, we obtain the energy barrier $E^*$ and the critical volume $V^*$ associated with nucleation as a function of  temperature; see Figs.~\ref{fig:energybarrier}g and h.
When the undercooling ($T_\text{eq} - T$) is small, the energy barrier $E^*$ for the semicoherent nucleus is lower than that of the coherent nucleus; this implies that the $\alpha$ plate will nucleate with semicoherent interfaces  along  the habit plane (see Fig.~\ref{fig:energybarrier}b).
However, when the undercooling is large, the coherent nucleus (see Fig.~\ref{fig:energybarrier}a) is favored at the incipient stage of nucleation.
Figure~\ref{fig:energybarrier}h shows that the critical volume of the semicoherent nucleus is smaller than that of the coherent nucleus at all temperatures.
Note, however, that  $\alpha$ often nucleates with the aid of $\omega$ precipitates~\cite{Nag_2009,rongpei_2019} or at grain boundaries of $\beta$~\cite{van2008nucleation}.
For such heterogeneous nucleation, the energy barrier and critical volume will be much lower than our prediction.
Nonetheless, the homogeneous nucleation predictions serve as a guideline for understanding heterogeneous nucleation effects.


\begin{figure}[tb]
\begin{center}
\includegraphics[width = 0.95\linewidth]{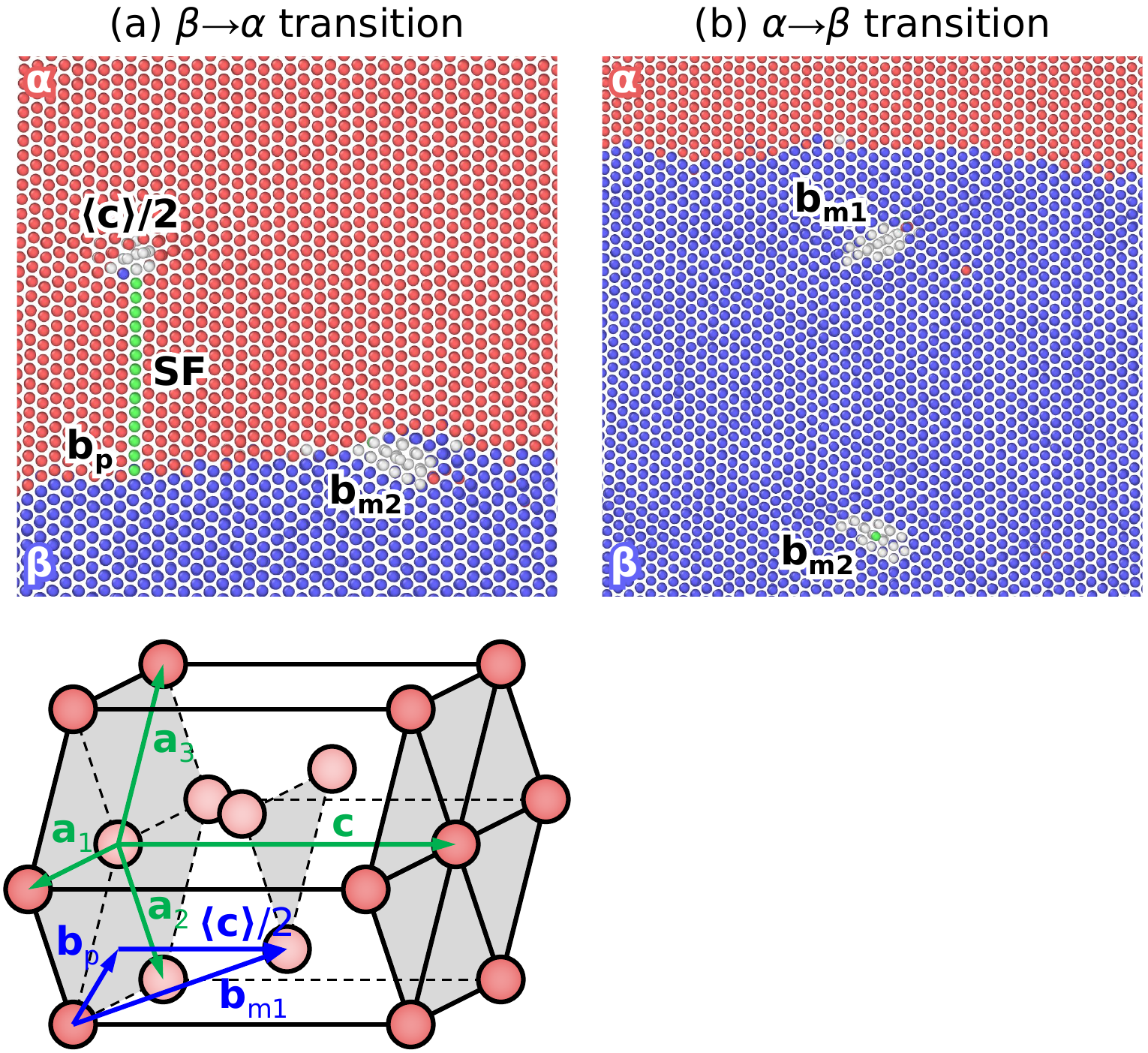}
\caption{Misfit dislocation migration behavior}%
{(a) When  $\alpha$ grows (downward interface migration), one BCC dislocation $\mathbf{b}_{\txm 1}$  dissociates into a  $\langle\mathbf{c}\rangle$  and a  basal $\mathbf{b}_\txp$ partial separated by a stacking fault. The HCP unit cell is also shown to identify the relevant Burgers vectors.
(b) When $\beta$ grows (upward interface migration), two BCC dislocations are left behind in $\beta$  and glide with the interface.
Atoms are colored by CNA. }
\label{fig:migration}
\end{center}
\end{figure}

The equilibrium semicoherent interface structure features arrays of steps and misfit dislocations (both are disconnections).
The interface may migrate via the glide of the steps along the interface, accompanied by the motion of the  misfit dislocations (dissociated into glissile BCC dislocations).
One of the BCC dislocations ($\mathbf{b}_{\txm 1}$ in Fig.~\ref{fig:semicoherent_structure}e) is on the interface while the other ($\mathbf{b}_{\txm 2}$) remains within  $\beta$.
Interface migration requires the cooperative motion of the two BCC dislocations.
When the interface migrates towards the $\beta$ phase (i.e.,  $\alpha$  grows; Fig.~\ref{fig:migration}a), the $\mathbf{b}_{\txm 1}$ dissociates as
\begin{equation}\label{bp1bbpc}
\begin{array}{ccccc}
\mathbf{b}_{\txm 1} &
\to&
\mathbf{b}_\txp&
+&
\langle\mathbf{c}\rangle/2 \\
\displaystyle{
\left[\bar{2}023\right] \frac{a_0^\alpha}{6} } &
\to&
\displaystyle{
\left[\bar{1}010\right]\frac{a_0^\alpha}{3} } &
+&
\displaystyle{
\left[0001\right]\frac{c_0^\alpha}{2} }
\end{array},
\end{equation}
where $\mathbf{b}_\txp$ and $\langle\mathbf{c}\rangle/2$ are the Burgers vectors of the partial dislocation on the basal plane and the $\langle\mathbf{c}\rangle$ edge dislocation in $\alpha$, respectively (Fig.~\ref{fig:migration}a).
The two dislocations resulting from the reaction in Eq.~\eqref{bp1bbpc} are separated by a stacking fault (green atoms in Fig.~\ref{fig:migration}a).
Experiments~\cite{lu2021structural,tan2016revealing,sarkar2014fcc} often show fine FCC lamellae and/or stacking faults within $\alpha$ or near the $\alpha$/$\beta$ interface.
The formation of these FCC lamellae and stacking faults may originate from the dissociation of misfit dislocations accompanying the interface migration associated  with $\alpha$ growth; in other words, the FCC lamellae and stacking faults terminated at the interface help to accommodate the mismatch along the interface.
The dissociation (Eq.~\eqref{bp1bbpc}) leads to the formation of a stacking fault; increase of the stacking fault area (and thus energy) during interface migration retards interface migration.
When the interface migrates towards  $\alpha$ ($\beta$ growth; Fig.~\ref{fig:migration}b), the two BCC dislocations, $\mathbf{b}_{\txm 1}$ and $\mathbf{b}_{\txm 2}$, will be left behind in $\beta$.
To accommodate the misfit along the $\alpha$/$\beta$ interface, $\mathbf{b}_{\txm 1}$ and $\mathbf{b}_{\txm 2}$ glide to follow the interface migration; again retarding interface migration. 
The similar phenomenon, i.e., stacking fault formation with interface migration in one direction, was also found for tilt grain boundaries in FCC metals~\cite{mccarthy2020shuffling}.
It was suggested that such direction-dependent stacking-fault formation could lead to the directionally anisotropic interface mobility. 

The above analysis shows that the introduction of steps eases interface migration while the introduction of misfit dislocations hinders it.
Lattice constant manipulation, for example, by alloying or straining, may be used to increase/decrease step and misfit dislocation density.
This may be employed to tune the $\alpha/\beta$ interface mobility and microstructure (e.g., lamella thickness).


To sum up, in this paper, we investigated the structure and thermodynamics of the $\alpha$/$\beta$ interface in Ti using molecular dynamics, thermodynamic integration and a DFT-trained Deep Potential.
Our major findings are as follows.
\begin{itemize}
\item[(i)]
The coherent interface free energy depends upon coherency strain (see Figs.~\ref{fig:fixedstrain} and \ref{fig:efenergy}).

\item[(ii)]
The structure of an equilibrium semicoherent interface consists of (a) an array of steps with step height $h_\txs = 2\sqrt{6}a_0^\beta /3$ and Burgers vector $b_\txs \approx 0.088 h_\txs$, (b) an array of misfit dislocations in the form of a pair of full dislocations in $\beta$  (BCC crystal), and (c) terraces with the structure of the coherent interface.
The intersection between a step line and a full dislocation line results in a local hump in the interface profile and bowing of the step.

\item[(iii)]
The equilibrium semicoherent interface free energy is 0.188~J~m$^{-2}$ (at the equilibrium $\alpha$/$\beta$ coexistence temperature, 1194~K).
The energy associated with the misfit dislocation array (lattice mismatch along the $[0001]^\alpha$ or $[110]^\beta$ direction), dominates the semicoherent interface free energy ($\sim 88$\%); the contribution associated with the coherent terraces is small ($\sim8.6$\%)

\item[(iv)]
The computed coherent/semicoherent interface free energy was used to predict the energy barrier, critical size and critical shape of the $\alpha$ nucleus (with coherent and semicoherent interfaces) within a $\beta$ matrix.
At large undercooling ($\Delta T \gtrsim 300$~K), the $\alpha$ precipitate  nucleates with coherent interfaces, which become semicoherent as the precipitate grows.
When the undercooling is small, the $\alpha$ precipitate forms and grows with semicoherent interfaces.

\item[(v)]
Analysis of the semicoherent interface structure shows that while the step array aids interface migration, the misfit dislocation array hinders it.
The misfit dislocation drag mechanism differs depending upon the direction of interface migration.
\end{itemize}


\section{Methods}\label{theory}

\subsection{Interface free energy calculation}

The free energy of an atomic system can be obtained by  $\lambda$ integration based on a set of thermodynamic equilibrium states.
$\lambda$ is a  parameter which is used to smoothly vary the Hamiltonian between a reference system $H_0$ and the Hamiltonian of the target system (i.e., the system of concern) $H_1$.
The ``mixed'' Hamiltonian is
$H(\lambda) = (1-\lambda)H_0 + \lambda H_1$.
The free energy of the target system, $F_1$, is obtained by  integration:
\begin{equation}\label{FFint}
F_1
= F_0 + \int_0^1 \left\langle
\frac{\partial H}{\partial \lambda}
\right\rangle_{\lambda} \ud \lambda,
\end{equation}
where $F_0$ is the reference system free energy and $\langle \cdot \rangle_\lambda$ is the ensemble average for a system with parameter $\lambda$.


We can perform a nonequilibrium MD simulation in which $\lambda$ changes with time $\lambda(t)$.
If the rate of  change of $\lambda$ is infinitesimally small, we  obtain the equilibrium state at each $\lambda$, the exact ensemble average $\langle \cdot \rangle_\lambda$ and an accurate evaluation of the integral in Eq.~\eqref{FFint}.
However, infinitesimally slow changes of $\lambda$  requires infinite  simulation cost.
We can construct a switching function $\lambda(t)$ which changes $\lambda$ from $0$ to $1$ (forward) and then from $1$ to $0$ (backward) with time.
The value for the integral Eq.~\eqref{FFint} is the difference between the work done in the forward  and backward processes~\cite{frenkel2001understanding,watanabe1990direct,de1996einstein,de1997adiabatic}:
\begin{equation}\label{Fadiabaticswitching}
F_1
= F_0 + \frac{1}{2}\left(
\overline{W^{0\to 1}_{\rm irr}}
- \overline{W^{1\to 0}_{\rm irr}}
\right),
\end{equation}
where the average irreversible work associated with switching $\lambda$  from $\lambda_1$ to $\lambda_2$ is
\begin{equation}
\overline{W^{\lambda_1 \to \lambda_2}_{\rm irr}}
= \int_{t(\lambda_1)}^{t(\lambda_2)}
\frac{\partial H}{\partial\lambda} \frac{\ud\lambda}{\ud t} \ud t
\end{equation}
 and $t(\lambda)$ is the inverse of $\lambda(t)$.


The choice of reference system  for determining the free energy in  $\lambda$ integration is important.
We choose a reference system for which we can easily determine the entropy (count the number of states).
We choose the Einstein crystal (EC), where each atom is an independent 3D harmonic oscillator, since only the vibrational entropy contributes to the free energy~\cite{frenkel1984new}.
The $N$-identical atom EC Hamiltonian is
\begin{equation}
H_0(\mathbf{r},\mathbf{p})
= \sum_{i=1}^{3N} \left(
\frac{p_i^2}{2m} + \frac{1}{2}m \omega^2 r_i^2
\right),
\end{equation}
where $\mathbf{r}$ and $\mathbf{p}$ are  generalized coordinates and momenta of atoms, $m$ is the mass and $\omega$ is the oscillator frequency.
The partition function and free energy are
\begin{align}
Z_0
&= \int
\exp\left(
-\frac{H_0(\mathbf{r},\mathbf{p})}{k_\uB T}
\right) \frac{\ud \mathbf{r} \ud \mathbf{p}}{h^{3N}}
= \left(\frac{k_\uB T}{\hbar \omega}\right)^{3N}, \\
F_0
&= -k_\uB T \ln Z_0
= 3Nk_\uB T
\ln \left( \frac{\hbar \omega}{k_\uB T} \right), \label{Ft3NkB}
\end{align}
where $k_\uB$ and $h$ are the Boltzmann and Planck constants and $\hbar=h/2\pi$.


With Eq.~\eqref{Ft3NkB}, Eq.~\eqref{Fadiabaticswitching} can be written as
\begin{align}
F_1(N,V,T)
&= 3 N k_\uB T \ln \left ( \frac{\hbar \omega}{k_\uB T} \right)
+ \frac{1}{2}\left(
\overline{W^{0\to 1}_{\rm irr}}
- \overline{W^{1\to 0}_{\rm irr}}
\right)
\nonumber\\
&+ k_\uB T \ln \left[
\frac{N}{V}
\left( \frac{2 \pi k_\uB T }{N m \omega^2} \right)^{3/2}
\right],
\end{align}
where the last term corrects  for the fixed center of mass~\cite{2021}.
Freitas et al.~\cite{Freitas_2016}  implemented  $\lambda$ integration with adiabatic switching in the Large-scale Atomic/Molecular Massively Parallel Simulator (LAMMPS)~\cite{plimpton1995fast}; we employ this here.


The interface free energy is the excess free energy of the system due to the presence of an interface separating two phases; the excess free energy is found by subtracting the bulk free energies of the two phases from the free energy of the two-phase system.
Frolov et al.~\cite{Frolov_2012} writes the interface free energy in crystalline materials as
\begin{equation}\label{gammaA}
\gamma A
= F - (\zeta^{\alpha}F^{\alpha} + \zeta^{\beta}F^{\beta}),
\end{equation}
where $\gamma$ is the interface free energy and $A$ is the interface area in the two-phase system; $F$, $F^\alpha$ and $F^\beta$ are the free energies of the two-phase system, and in  bulk $\alpha$  and  $\beta$.
$\zeta^\alpha$ and $\zeta^\beta$ are the fractions of two phases:
\begin{equation}
\zeta^{\alpha}
= \frac{NV^{\beta} - N^{\beta}V}
{N^{\alpha}V^{\beta} - N^{\beta}V^{\alpha}}, \quad
\zeta^{\beta}
= \frac{N^{\alpha}V - NV^{\alpha}}
{N^{\alpha}V^{\beta} - N^{\beta}V^{\alpha}},
\end{equation}
where $(N,V)$, $(N^\alpha,V^\alpha)$ and $(N^\beta,V^\beta)$ are the numbers of atoms and volumes in the two-phase system and  bulk $\alpha$ and $\beta$.
In this way, identification of the phases to which each atom belongs is unnecessary.

\subsection{Free-end nudged-elastic-band method}

The free-end nudged-elastic-band (FE-NEB) method is applied to find the minimum energy path and saddle points in Sec. \ref{section:discussion}. The NEB force on the $i^\text{th}$ image (except the end image) is
\begin{equation}
	\mathbf{F}_{i}^{\text{NEB}} = \mathbf{F}_{i}^{\perp} + \mathbf{F}_{i}^{\text{S}}. 
\end{equation}
$\mathbf{F}_{i}^{\perp}$ is the true force projected along the string normal.
$\mathbf{F}_{i}^{\mathrm{S}}$ is the spring force:
\begin{equation}\label{spring}
	\mathbf{F}_{i}^{\text{S}} 
	= k \left(
	\left| \mathbf{R}_{i+1} - \mathbf{R}_{i} \right| 
	- \left| \mathbf{R}_{i} - \mathbf{R}_{i-1} \right|
	\right) 
	\boldsymbol{\tau}_{i},
\end{equation}
where $\mathbf{R}_i$ is the configuration of the $i^\text{th}$ image, $\boldsymbol{\tau}_{i}$ is normalized local tangent at the $i^\text{th}$ image, and $k$ is a spring constant. We set $k$ = 50 \AA$^{-1}$ to obtain smooth NEB paths. 
The minima were obtained using the Quick-Min (QM) algorithm~\cite{sheppard2008optimization} with step size 0.005 and a stop criteria of $ \vert \Delta \mathbf{F}^\text{NEB}_{\max}\vert < 1.11\times10^{-8}$ (dimensionless).
The detailed algorithm of FE-NEB is given in SI. 

\subsection{Computational settings}

All the MD simulations were performed using LAMMPS~\cite{plimpton1995fast}.
Interactions between Ti atoms were described using a Deep Potential (DP)~\cite{wen2021specialising}. The DP for Ti predictions for basic properties of HCP, BCC and FCC Ti such as lattice parameters, cohesive energies, elastic constants, and defect structures/energies as well as defect properties (surface, point defect, stacking fault, $\gamma$-surface on multiple planes, dislocation core structures) and transformation and melting temperatures are shown in~\citet{wen2021specialising}. These are compared with experiment and/or DFT calculations where available. Overall, the agreement is excellent. This potential can also reproduce features of the thermal martensite transformation in Ti (see Supplementary Information). The potential and its properties are available from the Deep Potential library~\cite{dplib}.

\section{Data Availability}
Data supporting the findings of this study are available from the corresponding author (FD) on reasonable request.

\hspace{0.5cm}

\section{Competing interests}
The author declare no Competing Financial or Non-Financial Interests.

\section{Author Contributions}
S.W. performed the atomistic simulations. S.W. and J.H. analyzed the result. J.H. and D.J.S conceived and directed the project. T.W. developed the Deep Potential. All authors wrote the manuscript.

\section*{Acknowledgments}
\noindent SW and DJS gratefully acknowledges the support of the Hong Kong Research Grants Council Collaborative Research Fund C1005-19G. JH acknowledges support of the Early Career Scheme (ECS) grant from the Research Grants Council of Hong Kong SAR, China [Project No. CityU21213921] and Donation for Research Projects 9229061.

\bibliography{interfaceshort.bib}

\end{document}